\documentstyle[pbar2000,12pt]{article}
\newcommand{\lapproxe}{\stackrel{<}{\scriptstyle \sim}}
\newcommand{\gapproxe}{\stackrel{>}{\scriptstyle \sim}}
\newcommand{\gapproxeq}{\lower.7ex\hbox{$\;\stackrel{\textstyle>}{\sim}\;
$}}
\begin{document}
\title{ 
\vspace{-3cm}\begin{flushright} \small{hep-ph/0107016}
\\ \small{LA-UR-01-3521}  \end{flushright} \vspace{.5cm} 
Multi-GeV Gluonic Mesons}

\author{Philip R. Page     
      \\ {\em Theoretical Division, MS-B283,  Los Alamos National 
Laboratory} \\
      {\em Los Alamos, NM 87545}     
         }  
%

\maketitle

\begin{abstract}
Lattice QCD gives reliable predictions for hybrid charmonium and 
multi-GeV
glueball masses. Proton-antiproton annihilation may offer an excellent
opportunity for the first observation of these states. There are two 
distinct
possible programs: The search for $J^{PC}$-exotic {\it and} 
non-$J^{PC}$-exotic
states. The latter program represents substantially higher cross 
sections and
does not absolutely require partial wave analysis, two very attractive
features. The program can be performed with a varying $\bar{p}$ energy 
$<10$
GeV and a fixed target.
\end{abstract}

\section{Introduction \label{in}}

From the theoretical side, the most interesting problem in medium energy
physics is the mechanisms of strong QCD. Particularly, explicit glue 
tests
QCD's strong gluonic interactions. The spectroscopy of the new states of 
matter
with explicit glue, called {\it gluonic hadrons}, is essentially 
unknown.
Within the taxa of possible gluonic hadrons, charmonium hybrids are 
under
better theoretical control, due to heavy quarks, and experimentally 
cleaner,
due to smaller widths, than light flavour hybrids. Glueball masses are 
well
known theoretically, and there is the possibility that glueballs may be 
narrow
in experiment.

The kinematics of a $\bar{p}$ with energy $E$ colliding with a $p$ at 
rest
are expressed by
\begin{equation}\label{eq}
W^2 = 2m_p  \; (m_p+\sqrt{m_p^2+E^2}) \end{equation} with $W$ the
centre-of-mass energy and $m_p$ the mass of the proton. An immediate
application of this formula is that for $E<10$ GeV, $W < 4500$ MeV. As 
we shall
see, this is perfectly sufficient for the study of hybrid charmonia and
multi-GeV glueballs. The main qualitatively-new physics of interest 
above this
$W$ region are the lowest-mass experimentally-undiscovered baryons with 
new
flavours: double-charmed baryons. These will have $W \gapproxe 2 \times 
3650$
MeV, so that $E \gapproxe 27$  GeV. Also of interest are the lowest-mass
new-flavoured mesons: bottom-charm mesons. Here $W \geq 2 \times 
(6.40\pm
0.39\pm 0.13)$ GeV~\cite{pdg00} and $E \gapproxe 86$ GeV. It is clear 
that
there are vast energy deserts between $E$ of 10 GeV, 27 GeV and 86 GeV.

The competition for the physics program outlined here comes from the
proposed{\linebreak} Glue/Charm Factory at GSI Darmstadt, where $E< 15$ 
GeV
\cite{gsi}.

\section{Gluonic Mesons}

\subsection{Hybrid charmonia\label{hc}}

These are charm-anticharm-glue composites. Many quenched lattice-QCD 
mass
predictions for unmixed hybrid charmonium are available. Hence masses 
are under
theoretical control. The quantum numbers of the eight low-lying hybrids 
are
$J^{PC}=1^{--}$, $(0,1,2)^{-+}$, $1^{++}$, and $(0,1,2)^{+-}$.

Decays are under less control, although there is a fairly general 
selection
rule stating that the low-lying hybrid charmonia do not decay to 
$D\bar{D},\;
D^{\ast}\bar{D}^{\ast},\; D_s\bar{D}_s,\; D^{\ast}_s\bar{D}^{\ast}_s$ 
and have
only small decays to $D\bar{D}^{\ast},\; D_s\bar{D}^{\ast}_s$. These 
final
states are called {\it open charm}. The selection rule holds for 
OZI-allowed
decay with nonrelativistic quarks. The $q\bar{q}$ pair is assumed to be 
created
with nonrelativistic spin 1~\cite{page}. The phenomenology of meson 
decay
strongly supports spin 1 pair creation. The specific case of the
$J^{PC}= 0^{+-}$ hybrid is particularly interesting. Its decays to 
$D\bar{D},\;
D^{\ast}\bar{D}^{\ast},\; D^{\ast}\bar{D},\; D_s\bar{D}_s,\;
D_s\bar{D}^{\ast}_s,\; D^{\ast}_s\bar{D}^{\ast}_s$ are forbidden by 
general
principles due to quantum numbers. The threshold for one 
orbitally-excited and
one ground-state charmed meson, the $D^{\ast\ast}\!D$ threshold, is at 
4290
MeV. Below this threshold hybrid charmonium is narrow. Total widths for 
the
$0^{+-}$ and $1^{-+}$ states have, for example, been estimated as 
respectively
${\cal O}\; (5,\; 20)$ MeV below the threshold~\cite{dunietz,pagep}. The 
main
conclusion to draw is that there is a strong possibility that hybrid 
charmonia
are narrower than the $\sim 50$ MeV/$c^2$ widths of conventional 
charmonia in
the same mass region~\cite{pdg00}.

The most obvious search channel is the decay of hybrid charmonium to
conventional charmonium and light hadrons.
This final state signals the $c\bar{c}$ nature
of initial state. The most easily detected conventional charmonium is 
the
$\psi$ with a substantial branching ratio into dileptons~\cite{pdg00}.

In $p\bar{p}$ annihilation at LEAR there are indications that the 
light-flavour
hybrid-meson candidates $\hat{\rho}(1405)$ and $\rho(1450)$ are produced 
with
substantial cross sections \cite{wiedner}, comparable to or slightly 
less than
conventional mesons. This is the main production process studied to date 
where
hybrid mesons are significantly produced, underscoring the utility of
$p\bar{p}$ annihilation as a choice for hybrid production.

\subsection{Glueballs}

These are gluonic composites without quark content. Good quenched 
lattice-QCD
mass predictions are available for unmixed glueballs, providing 
theoretical
control.

Unmixed glueballs are believed to be narrow because the process whereby
$q\bar{q}$ pairs are created to enable decay into mesons is forbidden by 
the
OZI rule. I shall outline three further reasons to believe that 
multi-GeV
glueballs are narrow. Firstly, the flattening of the linear confining 
potential
between quarks due to pair creation at large $q\bar{q}$ separations 
implies
that there are no light-flavour conventional mesons $\gapproxe 3100\pm 
110$
MeV/$c^2$ in mass~\cite{bri}. Secondly, the mixing of glueballs with 
charmonium
is likely to be small due to the penalty incurred by the creation of a
$c\bar{c}$ pair. These arguments suggest that the glueball will not mix
substantially with light-flavour conventional mesons or charmonia. 
Thirdly, as
will be demonstrated below, most multi-GeV glueballs in the mass region 
of
interest do not decay to two other glueballs because of quantum numbers. 
The
above strongly argue that multi-GeV glueballs are narrow. The main fly 
in the
ointment is the possibility of mixing with light-flavour hybrid mesons.

The production of the glueball candidate $f_0(1500)$ in $p\bar{p}$
annihilation at LEAR is substantial, comparable to other mesons, but
weaker than, for example, the $f_2$ \cite{bugg}. Hence
$p\bar{p}$ annihilation is well suited for glueball production.
In fact, it is commonly thought to  be a ``glue-rich'' process.

The main focus of this paper will be on hybrid charmonium. The 
discussions of
hybrid charmonium and the glueball are somewhat different because hybrid
charmonium's telltale decay to $\psi$ and light hadrons is so different 
from
the purely light-hadron decays expected for the glueball. As we shall 
see,
there is also a large difference in production cross sections.

\section{$J^{PC}$ Exotics}

When a list of the possible $J^{PC}$ of conventional mesons is made, 
there are
certain $J^{PC}$'s which are not possible. These ``exotic'' $J^{PC} 
= 0^{--},\;
0^{+-},\; 1^{-+},\; 2^{+-},\ldots$ immediately indicate that the state 
is not a
conventional meson. In the mass region of interest, it is most likely to 
be
hybrid charmonium or a glueball, or possibly a four-quark state
($q\bar{q}q\bar{q}$). The detection of a $J^{PC}$ exotic state must 
hence be
assigned the highest priority in the search for gluonic mesons. The 
advantage
is that they cannot mix with conventional mesons. The flip side of the 
coin is
that the $J^{PC}$ must be established experimentally, {\it i.e.}\ one 
needs
detailed angular distributions and preferably a full partial-wave 
analysis
(PWA).

\subsection{Hybrid $c\bar{c}$: Mass}

Quenched lattice-QCD mass predictions for hybrid charmonia are given in
Table~\ref{tabQCDmass}.

\begin{table}[h!]
\begin{center}
\begin{tabular}{|c|c|r|}
\hline 
{ $1^{-+}$} & $0^{+-}$ & Ref.\\
\hline \hline 
{ $4410^{+60}_{-150}\pm\mbox{sys} $} & $4560^{+110}_{-100}\pm\mbox{sys} 
$ & \cite{ber1} \\
{ $4290^{+110}_{-190}\pm\mbox{sys} $} & $4560^{+80}_{-110}\pm\mbox{sys} 
$ & \cite{ber1} \\
{ $4390\pm 80\pm 200$} & $4610\pm 110\pm 200$ & \cite{ber2} \\
\hline 
\end{tabular}
\caption{\label{tabQCDmass} Quenched lattice-QCD mass predictions for 
hybrid charmonia in
MeV/$c^2$.}
\end{center}
\end{table}
As indicated in subsection \ref{hc}, below the $D^{\ast\ast}\!D$ 
threshold the
$1^{-+}$ and $0^{+-}$ hybrid charmonia are expected to be narrow. It is 
clear
that mass predictions for the $1^{-+}$ straddle the $D^{\ast\ast}D$ 
threshold,
while the $0^{+-}$ is most likely above the threshold. The $2^{+-}$ and
$0^{--}$ exotics are probably above the threshold too, if lattice 
calculations
for light-quark hybrids serve as a guide. Hence one concludes that there 
is
likely to be no more than one narrow exotic hybrid charmonium!

\subsection{Hybrid $c\bar{c}$: Production}

The first production process, $p\bar{p}{\rightarrow}$ exotic, is called 
{\it
formation}. This process can not produce $J^{PC}$-exotic states, though. 
This
follows because the $p\bar{p}$ system, just like the $q\bar{q}$ system, 
cannot
be $J^{PC}$ exotic. For $\bar{p}$ in flight a large tower of $J^{PC}$ is
accessed by the $p\bar{p}$ system, but none of these is exotic.

The second process, $p\bar{p}\rightarrow {\mathrm exotic} + (\pi^{0},\;
\pi\pi,\; \pi\pi\pi,\; \eta,\ldots)$, can produce $J^{PC}$-exotic 
states, and
is called {\it production}. When the extra light hadron, {\it e.g.}\ the
$\pi^{0}$, is accounted for, the condition $E < 10$ GeV is equivalent to
$m_{\mbox{exotic}} < 4360$ MeV/$c^2$. This bound is above the 
$D^{\ast\ast}D$
threshold, so that there is no need for a $\bar{p}$ energy above 10 GeV.

\subsection{Hybrid $c\bar{c}$: Decay}

\begin{table}[t]
\begin{center}
\begin{tabular}{|c|c|c|c|}
\hline
$J^{PC}$ & Open charm & Hidden charm & Light hadrons \\
\hline
\hline
$0^{+-}$ & Forbidden  & $J/\psi \{f_{\{0,1,2\}},(\pi\pi)_S\}$ &
 $a_{\{0,1,2\}} \rho;\; a_{\{1,2\}}\{b_1, \gamma\}$ \\
& for all & $h_c\eta;\; \eta_c h_1$&
 $b_1 \pi;\; h_1 \eta^{(')}$ \\
& cominations of  &
$\chi_{c0}\omega$&$\{(\pi\pi)_S, f_0\}\{\omega,\phi\}$\\
& $D^{(*)} D^{(*)}  $  &
$\chi_{c\{1,2\}}\{\omega,h_1,\gamma\}$&
$f_{\{1,2\}}\{\omega,h_1,\phi,\gamma\}$\\
\hline
$0^{--}$ & $D^* D$&
$h_c (\pi\pi)_S$&
$a_{\{0,1,2\}} b_1;\; a_{\{1,2\}}\{\rho, \gamma\}$ \\
   &   &$J/\psi\{f_{\{1,2\}},\eta^{(')}\} $&
$ \rho\pi$\\
    &           &
$\chi_{c0}h_1;\;\eta_c\{\omega,\phi\} $&
$f_0 h_1;\;\eta^{(')}\{\omega,\phi\}$\\
    &           &
$\chi_{c\{1,2\}}\{\omega,h_1,\gamma\} $&
$f_{\{1,2\}}\{\omega,h_1,\phi,\gamma\} $\\
\hline
$1^{-+}$ & $D^* D$,  $D^* D^*$ &
$\chi_{c\{0,1,2\}} (\pi\pi)_S$&
$a_{\{0,1,2\}} a_{\{0,1,2\}};\;a_{\{1,2\}}\pi$\\
         &          &
$\eta_c\{f_{\{1,2\}},\eta^{(')}\} $ &
$f_{\{0,1,2\}} f_{\{0,1,2\}};\;f_{\{1,2\}}\eta^{(')}$\\
         &           &
$\chi_{c\{1,2\}}\eta$&
$\{\rho,\gamma\}\{\rho,b_1\};\;b_1b_1$\\
    &           &$\{h_c,J/\psi\}\{\omega,h_1,\phi,\gamma\} $&
$\{\omega,h_1,\phi,\gamma\}\{\omega,h_1,\phi,\gamma\}$\\
\hline
$2^{+-}$ & $ D^* D$, $D^* D^*$     &
$\{h_c,J/\psi\} \{f_{\{0,1,2\}},(\pi\pi)_S\}$&
$a_{\{0,1,2\}}\{\rho,b_1,\gamma\} $
\\
& & $\{h_c,J/\psi\}\eta^{(')} $ &
$\{\rho,\gamma,b_1\}\pi $\\
& & $\{\eta_c,\chi_{c\{0,1,2\}}\} \{\omega,h_1,\phi,\gamma\} $ &
$\{\eta^{(')},f_{\{0,1,2\}}\} \{\omega,h_1,\phi,\gamma\}$\\
\hline
\hline
\end{tabular}
\caption{\label{tab29}
Some possible experimentally-accessible final states of $J^{PC}$-exotic 
hybrid
charmonia and glueballs below $D^{**}D$ threshold 
\protect\cite{dunietz}.
Decays to $p\bar{p}\{\pi,\eta^{(')},\omega,\rho,\phi\}$ are allowed for 
all
states listed.  }
\end{center}
\end{table}

From the list of possible decay modes in Table \ref{tab29} the easy ones
involving $\psi$ are $1^{-+} \rightarrow \psi\; (\omega,\phi,\gamma)
\rightarrow e^+ e^- e^+ e^-$ or $e^+ e^- \gamma$. The radiative decay is 
likely
to have a small branching ratio, since it is electromagnetic, and the 
$\omega$
and $\phi$ have small branching ratios to $e^+ e^-$~\cite{pdg00}. Direct
detection of all final-state particles may hence be problematic. In the 
decay
$0^{+-} \rightarrow \psi\; (\pi\pi)_S\;\; [\psi \pi^0\pi^0] \rightarrow 
e^+
e^-\gamma\gamma\gamma\gamma$ one looks at the $\pi^0\pi^0$ combination 
because
that can only be in an even wave by Bose symmetry. Here identification 
of all
final products is most likely to be hampered by the large $\pi\pi$ 
background
in $p\bar{p}$ annihilation. Instead of detecting all final-state 
particles, the
technique of missing mass may be more promising: detect only the $\pi^0$ 
and
$\psi$ in $p\bar{p}\rightarrow$ exotic $\pi^{0} \rightarrow \psi X 
\pi^{0}$.

\subsection{Hybrid $c\bar{c}$: Cross section}

The cross section for production of the $\psi$ is
$\sigma(p\bar{p}\rightarrow\psi\pi^0) = 130\pm 25$ pb~\cite{cester}. 
As pointed
out in subsection \ref{hc}, the $J^{PC}$-exotic light-flavour
$\hat{\rho}(1405)$ discovered at LEAR was observed in production at a 
similar
level to other light flavour mesons. Hence we shall take the production 
cross
section of hybrid charmonium to be 130 pb. With a luminosity of 
$10^{33}$
cm$^{-2}$ s$^{-1}$ foreseen for the new Fermilab $\bar{p}$ facility, 
$50\%$
efficiency, and a branching ratio $BR($exotic$\rightarrow
{\psi\omega}) = 1\%$, we estimate 10 $1^{-+}$ hybrid charmonia to be 
detected
via missing mass per day. This is not a promising rate, keeping in mind 
that
$\gapproxe 5000$ events were collected in the first 1$^{-+}$ light meson
discovery experiments at Brookhaven E852, due to the constraints imposed 
by
viable PWA.

It should be noted that considerable uncertainties exist in the 
assumptions used to derive the estimate of 10 hybrid charmonia
per day. Due to the small branching ratio 
$BR($exotic$\rightarrow {\psi\omega})$ 
used, the estimate should be regarded as conservative. 
Theoretical results indicate that 
$\sigma(p\bar{p}\rightarrow X\pi^0)$ is proportional to the width$(X\rightarrow
p\bar{p})$~\cite{maiani}. Because our estimate for hybrid charmonium
production was based on 
$\sigma(p\bar{p}\rightarrow \psi\pi^0)$, it is, according to ref.~\cite{maiani},
based on the width$(\psi\rightarrow p\bar{p})$. Experimentally, the
width$(X\rightarrow p\bar{p})$ ranges from $\sim 30\%$ of the value for the
$\psi$ (for $X=\psi(2S)$) to $\sim 90$ times the value for $\psi$
(for $X=\eta_c$). This considerable variation implies
large variation in  $\sigma(p\bar{p}\rightarrow X\pi^0)$
and hence substantial variation in the estimate of the amount of hybrid charmonia
produced.

\subsection{Glueballs \label{gl}}

The production of a glueball has a substantially larger cross section 
than
hybrid charmonium. This is easy to see: The light quarks comprising the
$p\bar{p}$ move into the outgoing light-flavour meson. Gluons are 
readily
converted to glueballs. For hybrid charmonium to be formed one requires, 
in
addition to the gluons, the costly creation of a $c\bar{c}$ pair.

A glueball, contrary to hybrid charmonium, is not expected to decay
to charmonium and light hadrons. This is because a glueball would have 
to
create a $c\bar{c}$ pair, while hybrid charmonium already has one 
present.

The glueball hence has the advantage over hybrid charmonium that its 
cross
section is large, but has the disadvantage that it has numerous decay 
channels
to light hadrons.

Quenched lattice QCD predicts an exotic $2^{+-}$ glueball at $4140\pm 
50\pm
200$ MeV/$c^2$~\cite{morn}. This is in fact the lightest exotic 
glueball. An
exotic $0^{+-}$ glueball is also predicted at $4740\pm 70\pm 230$
MeV/$c^2$~\cite{morn}, but it is too heavy relative to the 
hybrid-charmonium
masses of interest. Hence the search appears to be for just one 
glueball!

The $2^{+-}$ glueball can energetically decay to two of any of the 
$0^{++},\;
2^{++}$ and $0^{-+}$ glueballs. However, these energetically-uninhibited 
decays
are $C$-parity forbidden, thus not allowing the glueball to become wide.

\section{$J^{PC}$ Unknown}

Suppose the search for $J^{PC}$ exotics outlined in the previous section 
is
abandoned, due to low cross sections, the paucity of states (one hybrid
charmonium and one glueball), and the requirement of excellent angular 
coverage
and understanding of the detector imposed by PWA. There are several 
advantages
to abandoning the search for $J^{PC}$ exotics. Firstly, although PWA is 
always
preferable, it is possible not to do it and to resort to bump hunting. 
The
latter possibility will be our assumption for the remainder of this 
section.
When PWA is not performed, conventional charmonia in the mass region of
interest will also show up as bumps. Secondly, the narrowness or decay 
modes of
hybrid charmonia and glueballs are likely to be distinctive from 
conventional
charmonia, enabling discrimination. Lastly, small conventional 
charmonium
mixing with hybrid charmonium or a glueball is expected. The latter is 
due to
the penalty incurred by the creation of a $c\bar{c}$ pair, and the 
former is
due to the heaviness of the charm quarks which enable a Born-Oppenheimer
approximation, separating conventional and hybrid charmonia by virtue of 
their
orthogonal gluonic wave functions. (The preceding argues that the mixing 
matrix
elements are small. However, mixing can still be substantial in case of
coincidental mass degeneracies before mixing. Such a coincidence might 
in fact
occur for $1^{--}$ charmonia \cite{ger}.)

\subsection{Hybrid $c\bar{c}$: Mass\label{mh}}

As alluded to in subsection \ref{hc}, there are eight low-lying hybrid
charmonia. Three of these are exotic, which, as we shall see in the next
subsection, will not be of further interest. Quenched lattice QCD 
indicates
that the hybrid charmonia $1^{--}, (0,{1},2)^{-+}$ are less massive than
$1^{++}, ({0},1,{2})^{+-}$ \cite{man}.

\begin{table}
\begin{center}
\begin{tabular}{|c||c|c|}
\hline 
$J^{PC}$ & Gluon exchange \cite{dphil}  & Confinement \cite{merlin} \\
\hline \hline 
$0^{-+}$ &  $-180$           &  $\phantom{-}8$  \\
$1^{-+}$ &  $\phantom{\,}-\!50$ &  $\phantom{-}4$  \\
$1^{--}$ &  $\phantom{-2}60$ &  $\phantom{-}0$  \\
$2^{-+}$ &  $\phantom{-}210$ &  $-4$   \\
\hline 
\end{tabular}
\caption{The splittings between the four lowest-lying hybrid charmonia 
(in MeV).}
\end{center}
\end{table}

The splittings between the four lowest lying hybrid charmonia are 
indicated in
Table 3. The vector-gluon-exchange contribution was calculated in cavity 
QCD,
{\it i.e.}\ the spherical bag model,\footnote{This calculation is an
improvement of the calculation in Table 1 of ref.~\cite{close}. Here the
Z-topology and Coulomb diagrams~\protect\cite{dphil} were not included. 
Also,
an {\it ad hoc} value for the size of splitting was used.} setting the 
size of
splitting consistent with those observed between the $\psi(1S)$ and
$\eta_c(1S)$ and the $\psi(2S)$ and $\eta_c(2S)$. The scalar confinement
contribution was calculated from the Thomas precession in the flux-tube 
model,
and is clearly subdominant. The splittings are consistent with quenched 
lattice
QCD \cite{drum}. There are general arguments based on heavy-quark 
spin-orbit
splitting and the masses of light-flavour exotic hybrids from quenched 
lattice
QCD that suggest that $0^{-+} < 1^{-+} < 2^{-+}$ \cite{pagep}, 
consistent with
the above. One also obtains the following new prediction for the mass 
ordering:
$0^{-+} < 1^{-+} < 0^{+-} < 1^{+-} < 2^{+-}$ \cite{pagep}. We shall take 
the
$0^{-+}$ and $1^{--}$ hybrid charmonia as being most likely to be below 
the
$D^{\ast\ast}D$ threshold.

\subsection{Hybrid $c\bar{c}$: Formation}

The formation $p\bar{p}{\rightarrow}$ non-exotic is allowed. Because 
exotic
$J^{PC}$ cannot be formed, we shall only be interested in non-exotics in 
this
section. As was derived in section \ref{in}, $W<4500$~MeV. In order to 
access
different $W$, the $\bar{p}$ energy must be varied.

\subsection{Hybrid $c\bar{c}$: Decay}

In the formation $p\bar{p}{\rightarrow}$ non-exotic
$\rightarrow\psi X$ the $\psi$ is detected, and
$X$ is constructed from missing mass. Here
$X = \eta,\; \eta',\;\omega,\;\phi,\; \pi\pi,\; K\bar{K},\ldots$.
An interesting feature is that conventional charmonium is expected
to be suppressed relative to hybrid charmonium
in these final states, because conventional charmonium freely decay to
open charm, meaning that their branching ratios to the listed final 
states
are small.

Two of the simplest final states are $\psi\eta$ and $\psi\omega$. The 
former
can arise from the non-exotic hybrid charmonia $1^{--}$, $1^{+-}$, and 
the
latter from $0^{-+}$, $2^{-+}$, $1^{++}$.

Other possible final states are decays of non-exotic hybrid charmonium 
to $e^+
e^-$, $\mu^{+}\mu^{-}$, $\tau^{+}\tau^{-}$, and $\gamma\gamma$. However,
theoretically their widths are expected to be substantially smaller than 
for
conventional charmonium \cite{ono}.

\subsection{Hybrid $c\bar{c}$: Cross section}

We assume a formation cross section $\sigma = 0.1\; \mu$b for hybrid 
charmonium
\cite{close,kon}. This is smaller than the measured formation cross 
section of
the $\psi$. One should keep in mind that the cross section decreases 
strongly
with increasing $W$ \cite{zio}. With a luminosity of $10^{33}\ {\rm
cm}^{-2}{\rm s}^{-1}$, a $50\%$ efficiency, and {\em BR}$({\rm exotic}
\rightarrow{\psi\omega}) = 1\%$, we obtain 5000 events/day. This is a 
very
healthy rate.

Since no PWA is required, we can detect the decays,
non-exotic $\rightarrow \psi X$, for all $X$. This could well have
a branching ratio of $20\%$, yielding 100000 events per day.
However, as we shall see below, this procedure has the disadvantage of 
the
appearance of many overlapping states, making isolation of the
gluonic mesons difficult.

\subsection{Glueballs}

As in subsection \ref{gl}, glueball formation will be considerably 
enhanced
above that of hybrid charmonium, and glueballs will have many decay 
channels to
light hadrons. One expects the electromagnetic coupling of non-exotic 
glueballs
to $e^+ e^-$ $\mu^{+}\mu^{-}$, $\tau^{+}\tau^{-}$, and $\gamma\gamma$ to 
be
small. The glueballs predicted in the relevant mass region are the 
$1^{--}$ at
$3850 \pm 50\pm 190$~MeV/$c^2$, $2^{-+}$ at $3890\pm 40\pm 
190$~MeV/$c^2$,
$2^{--}$ at $3930\pm 40\pm 190$~MeV/$c^2$, and the $3^{--}$ at $4130\pm 
90\pm
200$~MeV/$c^2$~\cite{morn}.

The $1^{--}$, $2^{--}$, and $3^{--}$ glueballs cannot decay to two 
glueballs
for the same reason as in subsection \ref{gl}. The $2^{-+}$ glueball 
will have
P-wave decay to $0^{++}2^{++}$ glueballs and D-wave decay to 
$0^{++}0^{-+}$
glueballs, both near threshold. There is hence a distinct possibility 
that
decay to two glueballs will not allow the $2^{-+}$ to become wide.

\subsection{Bump hunting}

There is little reason to expect hybrid charmonium below 4 GeV. In the 
mass
region $4.0-4.3$ GeV/$c^2$, we expect 18 conventional charmonia: two 3S 
states
at $\sim 4.0$ GeV/$c^2$, four 1F states at $\sim 4.1$ GeV/$c^2$, four 2D 
and
four 1G states at $\sim 4.2$ GeV/$c^2$, and four 3P states at $\sim 4.3$
GeV/$c^2$ (the $D^{\ast\ast}D$ threshold) \cite{bri,torn}. Assuming that 
the
average resonance has a width of $\sim 50$ MeV/$c^2$ \cite{pdg00}, we 
expect
contiguous resonances in the mass region $4.0-4.3$ GeV/$c^2$ of 
interest. In
addition, we expect $\gapproxe 4$ non-exotic glueballs, {\it i.e.}
those
documented in the previous subsection, noting that not all glueballs in 
the
mass region of interest have probably been calculated by theory. There 
are also
$\lapproxe 5$ non-exotic hybrid charmonia, given that not all states 
documented
in subsection \ref{mh} will necessarily lie in the mass region of 
interest.

The contiguity of resonances is, however, much more conducive to
bump hunting when a specific decay channel is considered. For
example, only 3 of the 18
 conventional charmonia, 2 of the $\lapproxe 5$ hybrid charmonia,
and 2 of the $\gapproxe 4$ glueballs can decay in S- or P-wave to
$\psi\eta$. More gluonic than conventional mesons actually appear!
The seven resonances in the relevant mass region will clearly stand
out as bumps. Assuming a good understanding of conventional states,
the new states will be distinctive.

An energy scan in the $W=4.0 - 4.3$ GeV region in 10 MeV bins 
corresponds (from
Eq.\ \ref{eq}) to a $\bar{p}$ beam tuned to 50 MeV with 30 steps. 
Similarly, an
energy scan in 30 MeV bins corresponds to a $\bar{p}$ beam tuned to 150 
MeV
with 10 steps.

\section{Conclusions}

There are two distinct possible programs, the search for $J^{PC}$-exotic 
and
non-$J^{PC}$-exotic states as elaborated in Table \ref{jpcsearch}.
\begin{table}[!h]
\begin{center}\label{tab2}
\begin{tabular}{|c|c|}
\hline 
$J^{PC}$ exotics & $J^{PC}$ unknown \\
\hline 
Need PWA & Do not need PWA \\
Fixed $\bar{p}$ energy & Varying $\bar{p}$ energy\\
Low $\sigma$ & High $\sigma$ \\
\hline 
\end{tabular}
\caption{\label{jpcsearch}Possible search programs for $J^{PC}$-exotic 
and
non-$J^{PC}$-exotic states.}
\end{center}
\end{table}
A search for $J^{PC}$-exotic states would unambiguously isolate gluonic 
mesons.
However, the second program represents substantially higher cross 
sections and
does not absolutely require partial-wave analysis.

\vspace{.1cm}
I acknowledge Helmut Koch for making me aware of ref.~\cite{maiani}, as 
well as the workshop organizers for carefully reading the manuscript.

\end{document}